\documentclass[showpacs]{revtex4}
\usepackage{graphicx}
\usepackage{dcolumn}
\usepackage{amsmath}
\usepackage{color}
\definecolor{coolblack}{rgb}{0.0, 0.18, 0.39}
\definecolor{darkred}{rgb}{0.5,0,0}
\definecolor{darkgreen}{rgb}{0,0.5,0}
\definecolor{darkblue}{rgb}{0,0,0.5}
\definecolor{lapislazuli}{rgb}{0.15, 0.38, 0.61}
\definecolor{venetianred}{rgb}{0.78, 0.03, 0.08}
\definecolor{bleudefrance}{rgb}{0.19, 0.55, 0.91}
\definecolor{dogwoodrose}{rgb}{0.84, 0.09, 0.41}
\usepackage[linktocpage,colorlinks]{hyperref}
\hypersetup{colorlinks=true, citecolor=darkgreen, linkcolor=blue,urlcolor = darkblue}
\makeatletter
\def\btt#1{\texttt{\@backslashchar#1}}
\DeclareRobustCommand\bblash{\btt{\@backslashchar}} \makeatother

\begin{document}
\title{Deflection of Light by a Rotating Black Hole Surrounded by ``Quintessence"}
\author{Prateek Sharma $^{a}$}\email{prteeksh@gmail.com}
\author{Hemwati Nandan $^{a}$$^{b}$}\email{hnandan@associates.iucaa.in}
\author{Radouane Gannouji$^{c}$}\email{radouane.gannouji@pucv.cl}
\author{Rashmi Uniyal$^{d}$}\email{rashmiuniyal001@gmail.com}
\author{Amare Abebe$^{b}$}\email{amare.abbebe@gmail.com}
\affiliation{$^{a}$Department of Physics, Gurukula Kangri Vishwavidyalaya, Haridwar 249 404, Uttarakhand, India}
\affiliation{$^{b}$Center for Space Research, North-West University, Mahikeng 2745, South Africa}
\affiliation{$^{c}$Instituto de F\'isica, Pontificia Universidad Cat\'olica de Valpara\'iso, Casilla - 4950, Valpara\'iso, Chile}
\affiliation{$^{d}$Department of Physics, Government Degree College, Narendranagar-249 175, Uttarakhand, India}

\begin{abstract}
\noindent We present a detailed analysis of a rotating black hole surrounded by ``quintessence". This solution  represents a fluid with a constant equation of state, $w$, which can for example describe an effective warm dark matter fluid around a black hole. We clarify the conditions for the existence of such a solution and study its structure by analyzing the existence of horizons as well as the extremal case. We show that the deflection angle produced by the black hole depends on the parameters $(c,w)$ which need to obey the condition $cw<0$ because of the weak energy condition, where $c$ is an additional parameter describing the hair of the black hole. In this context, we found that for $w\simeq 0.1$ (consistent with warm dark matter) and $c<0$, the deviation angle is larger than that in the Kerr spacetime for direct and retrograde orbits. We also derive an exact solution in the case of $w=-1/3$.
\end{abstract}
\maketitle

\section{\bigskip INTRODUCTION}
The deflection of light by massive astrophysical objects remains one of the most interesting ways to study the geometry produced by these objects. We have recently completed the 100 years of the confirmation  of the bending of light ray by a star as observed by Eddington and Dyson, which is a weak effect, everyday used by astronomers to measure gravitational mass of clusters or galaxies. On the other side, we have also strong gravitational deflection of light, which occurs when a light ray approaches a massive object such as a black hole. If the photon comes too close to the object, it will not escape,  giving birth to a darker area known as the shadow of the black hole, which has been observed this year. It is an interesting effect which motivates the study of the geodesic equations and therefore the geometry of the spacetime.

The calculation of the bending angle \cite{Darwin,Atkinson,Luminet:1979nyg,Ohanian} for Schwarzschild and Kerr geometries show that
as we approach these objects, the bending angle exceeds $2\pi$, indicating that the particle could loop around the object multiple times. It is therefore important to systematically study this behavior for geometries different from the Kerr black hole which could become an interesting test of general relativity (GR).

In fact, the black holes in GR and alternative theories of gravity exhibit the largest curvature of spacetime accessible to direct or indirect measurements. They are,
therefore, ideal systems to test our theories of gravity under extreme gravitational conditions. Even if GR remains very successful, some questions remain without answers or at least with no general consensus. Among these questions is the dark matter \cite{Bertone:2016nfn} and the dark energy problems \cite{Huterer:2017buf} of late-time cosmology. As a way to study these models, a scalar field is often introduced \cite{Gannouji:2019mph} and among the multiple models, quintessence remains the simplest \cite{Ratra:1987rm,Caldwell:1997ii,Nandan:2016ksb,Uniyal:2014paa}. It is doubtful to think that dark energy could affect the surrounding of a black hole, but dark matter could produce an effective fluid around the compact object. Because dark matter does not interact with photons, it would not deflect light directly but the modification of the metric would affect the path of light rays (see e.g. \cite{Konoplya:2019sns} for the modification of the shadow of a black hole by dark matter). 

Cold dark matter has an equation of state (EOS) $w = 0$ like dust, while hot dark matter has an equation of state like radiation because it is relativistic matter with $w = 1/3$. A warm dark matter should have an effective equation of state with $0<w<1/3$. Also notice that some coupling of dark matter with other species might give an effective parameter of state different from zero. Recently, an analysis \cite{Kopp:2018zxp} in this direction gave $0<w<0.2$. 

Interestingly, a spacetime with an effective fluid described by an equation of state $P=w\rho$ exists \cite{Kiselev:2002dx} and has been extended to a rotating solution in  \cite{Toshmatov:2015npp}. Unfortunately, this solution is known in the literature as ``quintessence" while it does not represent a scalar field around a black hole but just a fluid with an equation of state $P=w\rho$. We will comment with more details on this nomenclature in Section II. 

One can notice a previous analysis \cite{Iftikhar:2019bhp} of circular orbits in the equatorial plane of this rotating black hole spacetime in the background of quintessential dark energy.
In the remaining sections of this paper, we will study the general description of spacetimes in Section II, followed by the study of the horizon and extremal solutions in Section III. In Section IV, we will numerically analyse the photon orbit and extend it further in Section V by studying the deflection angle for various parameters of the theory. Finally, in Section VI, we will find particular cases where the problem can be analysed exactly. We will conclude the main results of our study in Section VII.
\section{\bigskip Rotating Black Hole Surrounded by Quintessence}
The metric of a rotating black hole surrounded with quintessence \cite{Toshmatov:2015npp} is given by

\begin{equation}
\begin{aligned}
ds^{2}=-\left[1-\frac{\left(2 M r + c r^{1-3w} \right)}{\Sigma}\right]dt^{2} +\frac{\Sigma dr^{2}}{\Delta} \\ -2a sin^{2}\theta\left[\frac{2 M r + c r^{1-3w} }{\varSigma}\right] dtd\phi
+\Sigma d\theta^{2} \\ +sin^{2}\theta\left[r^{2} + a^{2} + a^{2}sin^{2}\theta\left(\frac{2 M r + c r^{1-3w}}{\varSigma}\right)\right] d\phi^{2}, \label{metric}
\end{aligned}
\end{equation} 
\noindent with
$\Delta = r^{2} + a^{2} -2Mr-cr^{1-3w}$ and $ \Sigma = r^{2}+a^{2}cos^{2}\theta $
where $c$ is a new parameter describing the hair of the black hole and
$w$ is quintessential equation of state parameter. First, we will check if this metric describes a solution of GR with quintessence field. For that we consider a fluid described as
\begin{align}
T_{\mu \nu}=\rho u_\mu u_\nu +P_r k_\mu k_\nu+P_t \Pi_{\mu \nu}\;,
\end{align}
where $\rho$ is the energy density of matter, $P_r$ is the radial pressure, $P_t$ the tangential pressure, $u_\mu$ is the normalized time-like fluid 4-velocity (such that $u_\mu u^\mu=-1$), $k_\mu$ is a unit space-like radial vector, $\Pi_{\mu \nu}=g_{\mu \nu}+u_\mu u_\nu -k_\mu k_\nu$ is the projection operator onto a 2-surface orthogonal to both vectors $u_\mu$ and $k_\mu$. We define also the angular velocity $\Omega=a/(a^2+r^2)$. Therefore, we have $u^3=\Omega u^0$ and from the condition of normalization, we obtain
\begin{align}
u^0=\Bigl[-(g_{00}+2\Omega g_{03}+\Omega^2 g_{33})\Bigr]^{-1/2}
\end{align}
and $k^1=(g_{11})^{-1/2}$.

From the expression of the metric, it is easy to calculate $T_{\mu \nu}=G_{\mu \nu}$, from which we obtain
\begin{align}
\rho&=T_{\mu \nu}u^\mu u^\nu\;,\\
P_r&=T_{\mu \nu}k^\mu k^\nu\;,\\
T &=-\rho+P_r+2P_t\;,
\end{align}
or more specifically
\begin{align}
\rho &= -3c w \frac{r^{1-3w}}{\Sigma^2}\;,\\
P_r &= 3c w \frac{r^{1-3w}}{\Sigma^2}\;,\\
P_t &= \rho-\frac{3cw(3w-1)r^{-1-3w}}{2\Sigma}\;.
\end{align}
We define the pressure as the average $3P=P_r+2P_t$, which therefore gives the equation of state
\begin{align}
\frac{P}{\rho}=w+(3w-1)a^2\frac{\cos^2\theta}{3 r^2}\;.
\end{align}
Notice that in the case $a=0$, we recover the quintessence solution \cite{Kiselev:2002dx}. We should mention that Kiselev \cite{Kiselev:2002dx} named this field ``quintessence"  as we do because the solution is known by this name in the literature even if it describes a fluid with a constant equation of state and not a quintessential one (see e.g. \cite{Visser:2019brz} for more details). Generically, the metric does not describe a quintessence field (we use the word quintessence as explained previously) but in the equatorial plane $\theta=\pi/2$, it reduces to $P=w\rho$. We will in the rest of the paper work in the equatorial plane to assure the right equation of state. Again, we would like to reinforce the idea that this solution does not represent a quintessence field, but a fluid with an equation of state $P=w\rho$ (as dark matter, baryonic matter or radiation) where the pressure is defined as the average pressure $3P=P_r+2P_t$.

Considering the equatorial plane, we will have
\begin{align}
\rho &= -\frac{3c w}{r^{3(1+w)}}\;,\\
P_r &= \frac{3c w}{r^{3(1+w)}}\;,\\
P_t &= -\frac{3c w(3w+1)}{2r^{3(1+w)}}\;.
\end{align}
Therefore, we need to impose the condition $cw<0$ to ensure $\rho>0$ and the weak energy condition, which in return implies $P_r<0$ and $P_t<0$ for $w<-1/3$.

\section{\bigskip Horizons}
\begin{figure}
	\centering
	\includegraphics[scale=0.88]{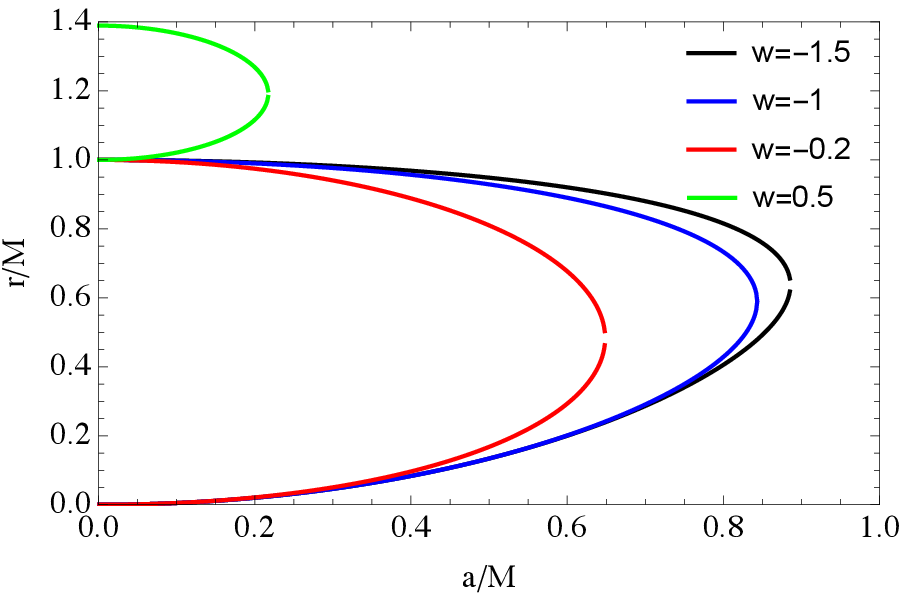}
	\includegraphics[scale=0.88]{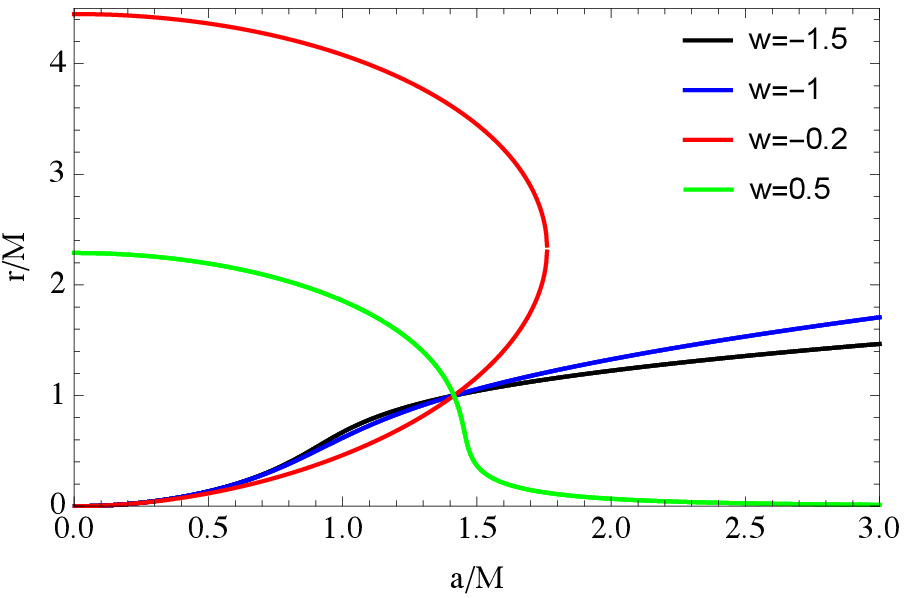}
	\caption{Variation of the position of the outer and inner horizon with the spin parameter for different values of $w$ and for $\alpha=-1$ in the upper panel and $\alpha=+1$ in the lower panel.}
	\label{fig:horizon}
\end{figure}

In this section, we will study the location of the horizons. They are defined by the equation $\Delta=0$ or more specifically
\begin{align}
r^2+a^2-2Mr-cr^{1-3w}=0\;.
\end{align}
The case $c\neq 0$ is complex and will be studied numerically. Considering normalized quantities such as $r\rightarrow r M$, $a\rightarrow a M$, $c \rightarrow \alpha M^{1 + 3 w}$, we obtain
\begin{align}
r^2+a^2-2r-\alpha r^{1-3w}=0\;.
\label{eq:HOR}
\end{align}
We see from Fig.(\ref{fig:horizon}) that for $c<0$, we have always 2 horizons, an outer or event horizon and an inner or Cauchy horizon, until some critical value of the spin parameter which defines the extremal black hole. In the case where $c>0$, the situation is more complicated. We notice the existence of 2 horizons for some values of $w$, we could have only 1 horizon and in some cases at least 3 horizons are possible.

\begin{figure}
	\centering
	\includegraphics[scale=0.6]{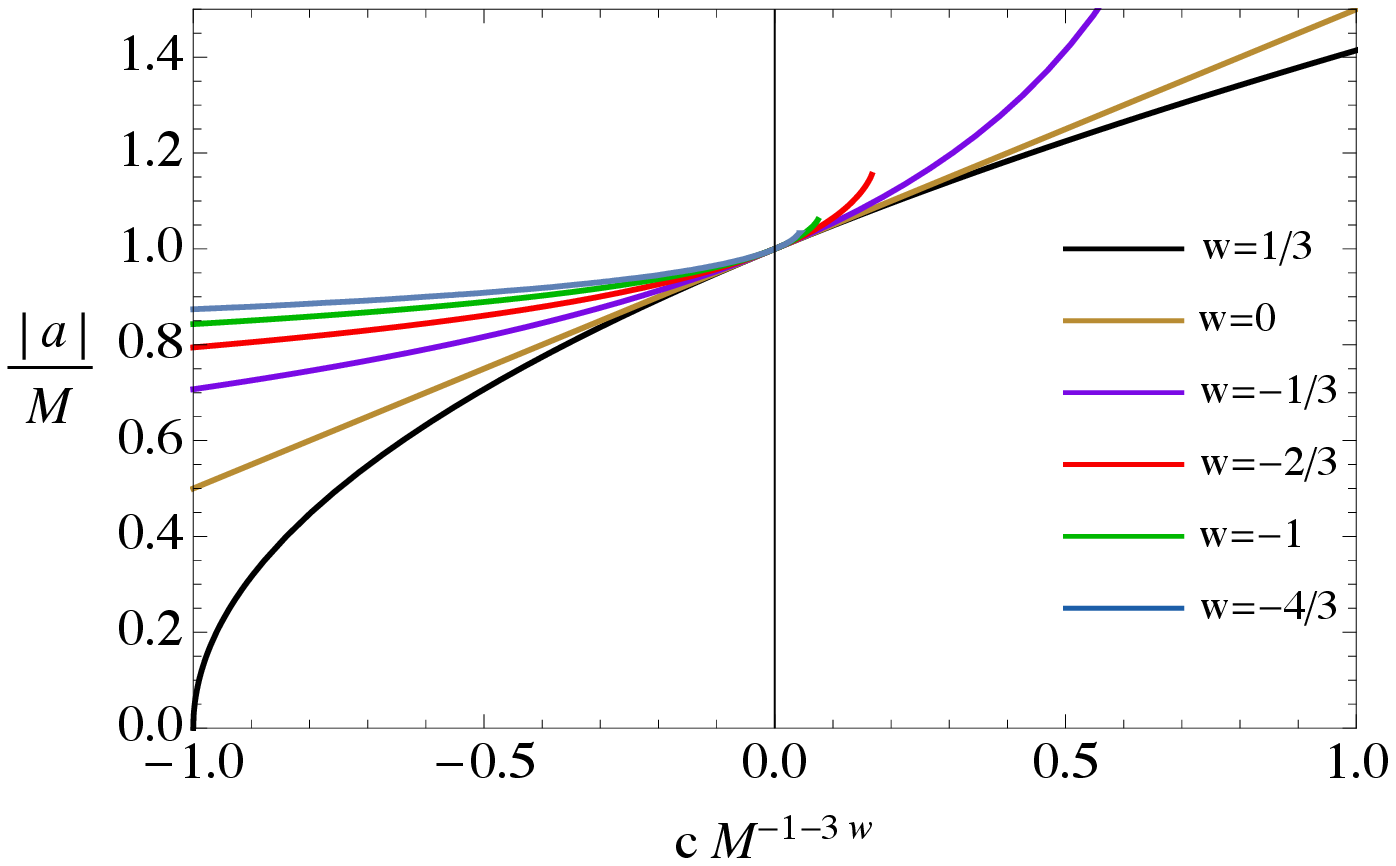}
	\includegraphics[scale=0.9]{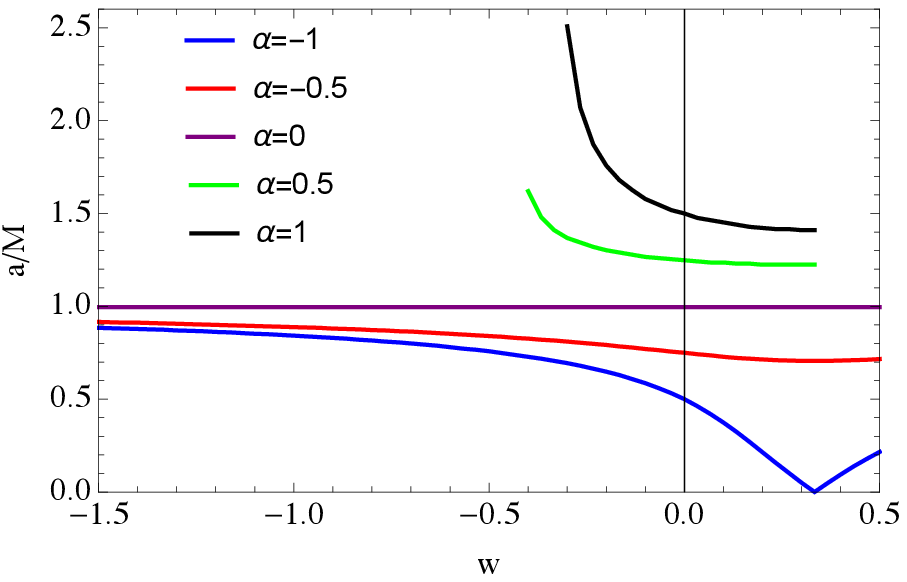}
	\caption{Variation of the extremal value of $|a|/M$ as a function of $cM^{-1-3w}$ (which corresponds to the new normalized parameter or hair of the theory) for various values of $w$ in the upper panel and as a function of $w$ for different values of $\alpha=c M^{-1-3 w}$ in the lower panel.}
	\label{fig:extremal}
\end{figure}

We see also in Fig.(\ref{fig:extremal}) the existence of a critical value defined by $w=1/3$. For $w<1/3$, we found that for any $c$ or $\alpha=c M^{-1-3 w}$, the spin parameter needed to reach an extremal black hole is smaller when $w$ increases. This behavior changes when $w>1/3$. Notice also that extremal black holes always exist for any $w$ when $\alpha<0$ while for $\alpha>0$, these black holes exist only in some range of $w$ which is always bounded from above by $w=1/3$. Finally, we notice that for $c<0$, the extremal black hole is smaller than the extremal Kerr while for $c>0$, the extremal solution is larger.

Finally, we notice from eq.(\ref{eq:HOR}) in the case of $a=0$, that $r=0$ is not a solution when $w>1/3$. Therefore the inner horizon, does not asymptote to $r=0$ when spinning parameter is zero. This can be seen in Fig.(\ref{fig:horizon}) for $w=0.5$.

\section{\bigskip Photon orbit}
The photon orbit is an interesting observable which is related to the black-hole shadow recently observed \cite{Akiyama:2019cqa}. Considering the orbit of photons in the previous spacetime, one can obtain the geodesic equations and principally the radial equation, written as $\dot r^2=f(r)$, where $f(r)$ is a function of $r$. The formalism is well known and we will briefly comment on the steps to take to obtain the final result. Considering the Lagrangian defined by $L=g_{\mu\nu}\dot x^\mu \dot x^\nu/2$ from which we deduce the generalized momenta $p_\mu\equiv \partial L/\partial x^\mu$, we can define the Hamiltonian $H=p_\mu \dot x^\mu-L$. Finally, using $p_t=-E=\text{constant}$ and $p_\phi=L=\text{constant}$ and $\theta=\pi/2$, it is easy to find $H=H(r,\dot r^2)$. Solving the equation $H=0$ (for photons), we find 
\begin{align}
\frac{\dot r^2}{E^2}\equiv f(r)=1+\frac{2M(a-d)^2}{r^3}+\frac{a^2-d^2}{r^2}+c\frac{(a-d)^2}{r^{3(1+w)}}
\label{eq:orbit}
\end{align}
where we defined $d=L/E$.

\begin{figure}
	\centering
	\includegraphics[scale=0.9]{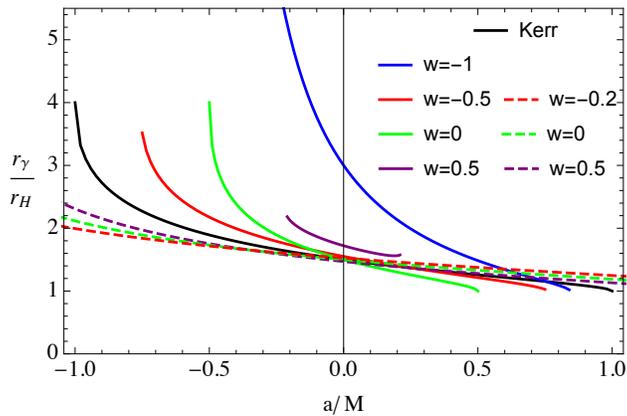}
	\caption{Variation of the photon orbit radius over the event horizon as a function of the spinning parameter $a/M$. In black, we have represented the evolution for the Kerr black hole, in plain line, the quintessence black hole for $\alpha=-1$ and finally in dashed line, we have $\alpha=1$ for various values of $w$.}
	\label{fig:photon}
\end{figure}

From eq.(\ref{eq:orbit}) we can find easily the equation for circular orbits, $f(r)=0,~f'(r)=0$ for which we have checked numerically that $f''(r)>0$. Therefore this orbit corresponds to the largest unstable circular orbit, viz. the photon sphere \cite{H.nandan2018}. For Kerr black hole, the photon orbit is $3M$ in the absence of rotation, and in the extremal case we have $M$ for direct orbits and $4M$ for retrograde orbits, which correspond to the smallest possible radii of the circular photon orbits. In the case of quintessence rotating black hole, we have a richer structure due to the 2 additional parameters $(c,\omega)$. Instead of studying only the photon orbit, it is more interesting to analyse the photon orbit ($r_\gamma$) over the event horizon radius $(r_H)$, namely $r_\gamma/r_H$. In fact, in the Kerr black hole, it is then easier to see that the photon orbit is getting closer to the event horizon for direct orbit when the spinning increases, until the extremal Kerr black hole, when the photon orbit is equal to the event horizon. We see in Fig. (\ref{fig:photon}) that $w=-1$ and $\alpha=-1$ which corresponds to Kerr-AdS is very different but well known. Focusing on other cases, we see that for a given retrograde orbit at fixed spin of the black hole $a$, the photon orbit is smaller for $c>0$ and larger for $c<0$ which is expected as we have an additional term in the metric which introduces an additional ``force" proportional to $(1+w)c$. Considering only normal quintessence, $w>-1$, we have for $c>0$ an additional attraction competing against the spinning while for $c<0$ there is an additional repulsion. This effect is also affected by the value of $w$. Finally, for direct orbits of the photons, the behavior is very similar for most of the parameters taking into account, as we have seen previously, that we can reach the extremal black hole for much smaller values of $a$ when $c<0$.

\section{\bigskip Deflection angle}

We consider a light ray that starts at infinity and approaches the black hole until some distance $r_0$, namely the distance of closest approach. It then emerges and arrives to an observer who is in some asymptotic region. We can calculate the change in the coordinate $\phi$ or equivalently the deflection of the light ray from its original path, $\theta$. Both are easily related by $\theta=\phi-\pi$. The point of closest approach is obtained by considering the largest root of the equation $\dot r^2=0$ and $\phi$ is defined by
\begin{align}
\phi=2\int_{r_0}^\infty \frac{{\rm d}\phi}{{\rm d}r} {\rm d}r\;.
\end{align}
The factor $2$ takes into account when the photon gets closer to the black hole until $r_0$ and moves away from $r_0$ to the asymptotic region. It is customary to represent it as a function of a normalized impact parameter, $b$, which is defined from the critical impact parameter, $d_c$, as 
\begin{align}
b=1-\frac{d_c}{d}
\end{align}
where $d$ is the impact parameter defined as $d=L/E$. The critical impact parameter is the smallest value of $d$ which produces a change of direction of the trajectory. For example, $d_c=3\sqrt{3}M$ for Schwarzschild spacetime, $2M$ for direct orbits in the extremal Kerr of $7M$ for the same extremal Kerr spacetime for retrograde orbits. 

We see easily that when $b=0$, the impact parameter is equal to the critical value, implying a radius equal to the photon orbit; the photon will not emerge and therefore $\theta$ diverges. This corresponds to the strong deviation regime, while for $b=1$, the impact parameter is infinite and therefore the deviation is zero which corresponds to the weak deflection regime.

\begin{figure}
	\centering
	\includegraphics[scale=0.85]{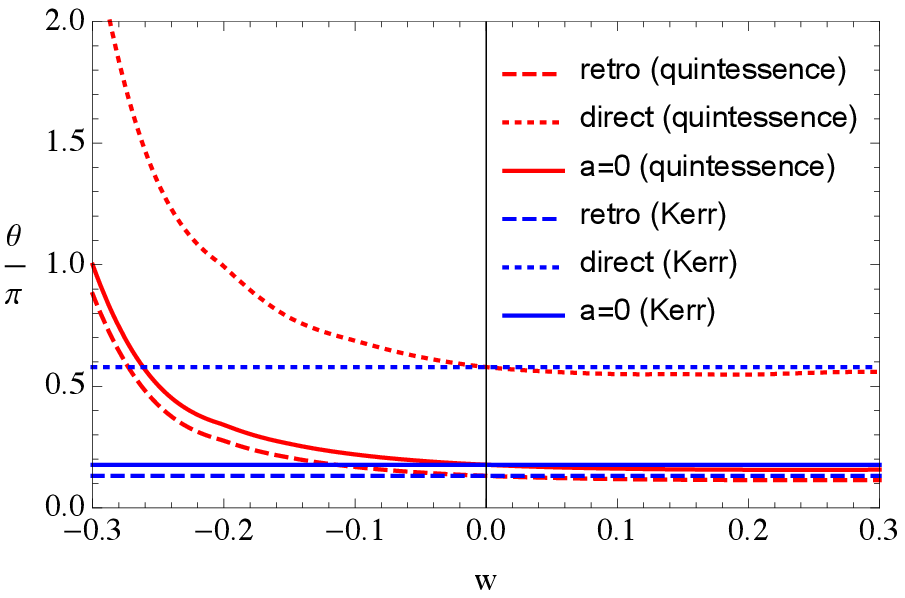}
	\includegraphics[scale=0.85]{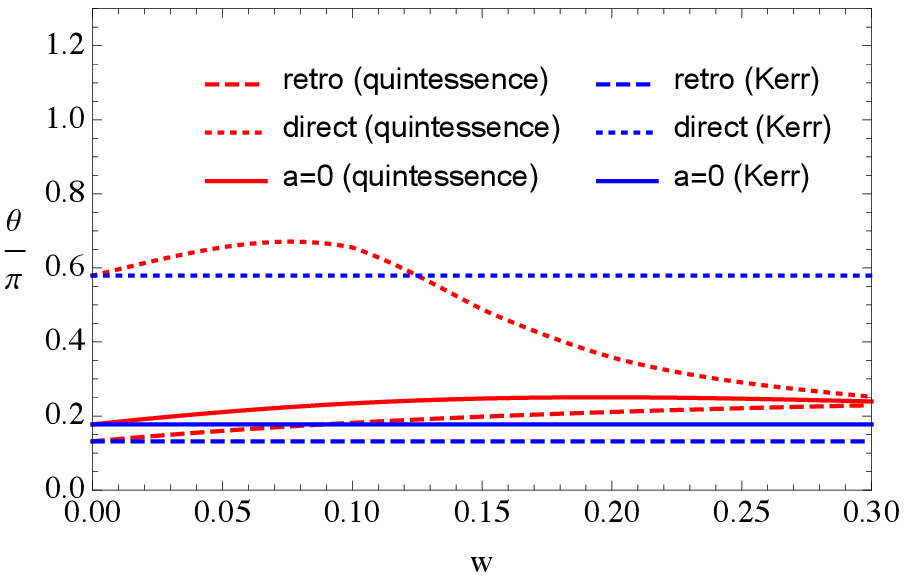}
	\caption{Variation of the bending angle for a normalized impact parameter $b=0.5$ as a function of $w$ and $\alpha=1$ in the upper panel and $\alpha=-1$ in the lower panel. The retrograde and the direct orbit are calculated for an extremal black hole.}
	\label{fig:bending1}
\end{figure}

We have represented in the Fig. (\ref{fig:bending1}) the deflection angle for $\alpha=1$. We have also represented the Kerr solution for comparison. Of course, both solutions produce the same deviation angle for $w=0$ (which corresponds to Kerr black hole with a different mass). The bending angle is always greater for direct orbits and it is greater than that for the Kerr spacetime for $w<0$. Notice that retrograde and direct orbits are calculated for extremal black holes.

\section{\bigskip Exact solutions}

As we have seen previously, the bending angle is defined as 
\begin{align}
\theta=2\int_{r_0}^\infty \frac{{\rm d}\phi}{{\rm d}r}{\rm d}r-\pi\;.
\end{align}
It is usually simpler to work with the variable $u=1/r$ which gives
\begin{align}
\theta=2\int_0^{\frac{1}{r_0}} \frac{f(u)}{\sqrt{B(u)}}{\rm d}u-\pi
\label{eq:theta}
\end{align}
where
\begin{align}
f(u)=\frac{b+2 (a - b) u+\alpha (a - b) u^{1 + 3 w}}{1-2u+a^2 u^2-\alpha u^{1+3w}}
\end{align}
and
\begin{align}
B(u)&=1+(a^2-b^2)u^2+2(a-b)^2u^3\\\nonumber
&+\alpha (a-b)^2u^{3(1+w)}\;.
\end{align}
Notice that as previously, we have normalized the variables $c \rightarrow \alpha M^{1 + 3 w}$, $L \rightarrow b E M$, $a \rightarrow a M$ and $r \rightarrow r M$.

\begin{figure}
	\centering
	\includegraphics[scale=0.64]{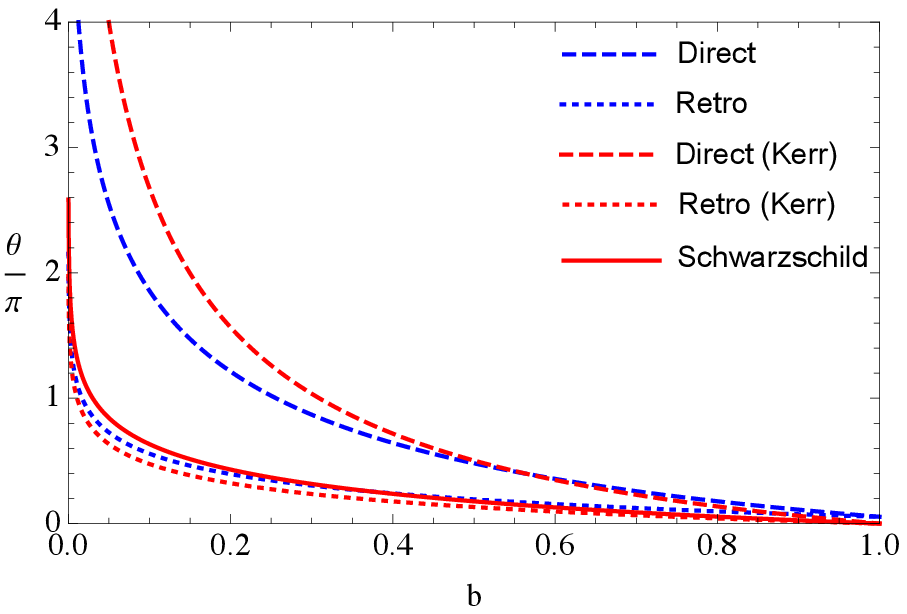}
	\includegraphics[scale=0.64]{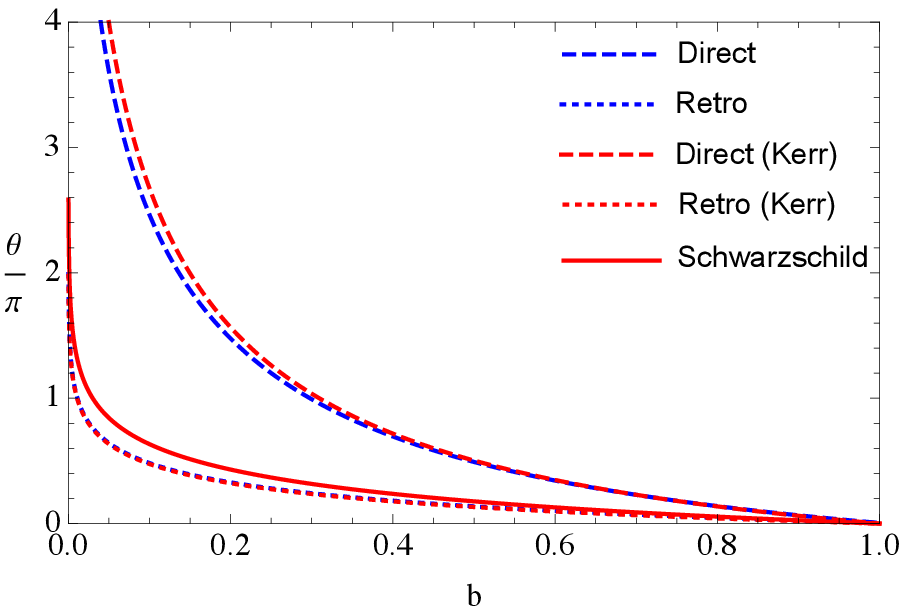}
	\caption{Variation of the bending angle of rotating quintessence for $|a|=0.99$, as a function of the normalized impact parameter $b$, for $w=-1/3$. The results are compared to Kerr spacetime for similar rotating parameter $|a|=0.99$, and Schwarzschild spacetime. In the left panel, we considered $\alpha=0.1$ and $\alpha=0.01$ in the right panel.}
	\label{fig:bendingexact}
\end{figure}

Focusing now on the particular case $w=-1/3$, we can decompose $f(u)$ in terms of partial fractions
\begin{align}
f(u)=\frac{a-b}{a^2}\Bigl[\frac{C}{u-u_+}+\frac{2-C}{u-u_-}\Bigr]
\end{align}
where 
\begin{align}
u_{\pm}=\frac{1 \pm \sqrt{1+a^2 (\alpha-1)}}{a^2}
\end{align}
and
\begin{align}
C=1+\frac{2(a-b)+a^2\Bigl[a\alpha+b(1-\alpha)\Bigr]}{2(a-b)\sqrt{1+a^2(\alpha-1)}}\;.
\end{align}
Here $B(u)$ is a cubic polynomial which has some particular characteristics, $B(u=0)=1>0$, $\lim_{u\rightarrow \infty }B=+\infty$. It is also easy to see that around $u=0$, $B$ is a decreasing function, therefore $B$ is first decreasing and then increasing. We conclude that it can have $0,1$ or $2$ real positive roots. We will name these roots $u_1<u_2\leq u_3$ which implies that
\begin{align}
B(u)=2(b-a)^2(u-u_1)(u-u_2)(u-u_3)\;.
\end{align}
For some parameters, only 1 real positive root exists. It corresponds to $B'(u)=0$ and $B(u)=0$ and therefore defines the critical impact parameter $b_c$
\begin{align}
&b_c^3+3a\frac{1+\alpha}{1-\alpha}b_c^2-3\frac{9-a^2(1-\alpha)(1+\alpha)^2}{(1-\alpha)^3}b_c\nonumber\\
&+a\frac{27+a^2(1+\alpha)^3}{(1-\alpha)^3}=0\;.
\end{align}
Performing the change of variable 
\begin{align}
b_c=\frac{6}{(1-\alpha)^{3/2}}\cos{x}-a\frac{1+\alpha}{1-\alpha}
\end{align}
we get $\cos{3x}+a\sqrt{1-\alpha}=0$
which implies
\begin{align}
b_c=\frac{6 \cos \left[\frac{1}{3} \cos ^{-1}\left(-a \sqrt{1-\alpha }\right)\right]}{(1-\alpha )^{3/2}}-a\frac{1+\alpha}{1-\alpha}\;.
\end{align}
In summary, we will need to consider $b>b_c$, for which we will always have two positive real roots of $B(u)$, namely, $u_2$ and $u_3$. Considering $u_2$ as the smallest root, it defines the position $r_0=1/u_2$ entering in the definition of the deviation angle (\ref{eq:theta}). Therefore, we have
\begin{align}
\theta &=-\frac{\sqrt{2}}{a^2}\Bigl[C\int_0^{u_2}\frac{{\rm d}u}{(u-u_+)\sqrt{(u-u_1)(u-u_2)(u-u_3)}}+\nonumber\\
&(2-C)\int_0^{u_2}\frac{{\rm d}u}{(u-u_-)\sqrt{(u-u_1)(u-u_2)(u-u_3)}}
\Bigr] -\pi\;.
\label{eq:bend}
\end{align}
These integrals can be written in terms of complete and incomplete elliptic integrals of the third kind respectively,

\begin{widetext}
	\begin{align}
	&\int_0^{u_2}\frac{{\rm d}u}{(u-u_\pm)\sqrt{(u-u_1)(u-u_2)(u-u_3)}}=-\int_{u_1}^0\frac{{\rm d}u}{(u-u_\pm)\sqrt{(u-u_1)(u_2-u)(u_3-u)}}\nonumber\\
	&\qquad\qquad\qquad\qquad \qquad\qquad \qquad\qquad \qquad\qquad \qquad\qquad  +\int_{u_1}^{u_2}\frac{{\rm d}u}{(u-u_\pm)\sqrt{(u-u_1)(u_2-u)(u_3-u)}}\nonumber\\
	&=\frac{2}{(u_\pm-u_1)\sqrt{u_3-u_1}}\Bigl[\int_0^{p}\frac{{\rm d}\theta}{(1-\frac{u_2-u_1}{u_\pm-u_1}\sin^2{\theta})\sqrt{1-\frac{u_2-u_1}{u_3-u_1}\sin^2{\theta}}}-\int_0^{\pi/2}\frac{{\rm d}\theta}{(1-\frac{u_2-u_1}{u_\pm-u_1}\sin^2{\theta})\sqrt{1-\frac{u_2-u_1}{u_3-u_1}\sin^2{\theta}}}\Bigr]\nonumber\\
	&=\frac{2}{(u_\pm-u_1)\sqrt{u_3-u_1}}\Bigl[\Pi\Bigl(\frac{u_2-u_1}{u_\pm-u_1};p,\frac{u_2-u_1}{u_3-u_1}\Bigr)-\Pi\Bigl(\frac{u_2-u_1}{u_\pm-u_1};\frac{u_2-u_1}{u_3-u_1}\Bigr)\Bigr]
	\label{eq:specialF}
	\end{align}
\end{widetext}

where we have performed the change of variable $u=u_1+(u_2-u_1)\sin^2{\theta}$ and defined $p=\arcsin{\sqrt{-\frac{u_1}{u_2-u_1}}}$. Using eqs.(\ref{eq:specialF},\ref{eq:bend}), we get the final result of the exact bending angle of a photon for $w=-1/3$. In Fig.(\ref{fig:bendingexact}), we have represented the exact bending angle as a function of the normalized impact parameter, $b$. When $a>0$, we consider the trajectory as direct and when $a<0$, the trajectory of the photon is retrograde. We see that for a given normalized impact parameter $b<0.6$, the direct orbit in the case of rotating quintessence has a smaller deviation than Kerr, while the retrograde orbit is slightly larger than Kerr. It is interesting to notice that even if for some particular parameter $b$, a rotating quintessence is similar to Kerr; the study of these spacetimes for various impact parameters permits to distinguish them.

\section{\bigskip Conclusions}                                 
We have studied the rotating quintessence black holes in the equatorial plane. We have analysed the structure of the solutions, the existence of extremal black holes and clarified if these spacetimes describe rotating black holes surrounded by a quintessence field. In that direction, we found that an extremal black hole always exists for $c<0$. We noticed that for $c <0$, the extremal black hole is smaller than the extremal Kerr while for $c >0$, the extremal solution is larger than in the Kerr spacetime. We investigated the photon orbit for the model. We found that for a given retrograde orbit, the photon orbit is smaller for $c >0$ and larger for $c <0$, as compared to Kerr black hole. Then we studied the deflection angle in the equatorial plane. We found that for $c>0$ and $w<0$, which does not violate the weak energy condition, the quintessential rotating black hole produces a larger deviation than Kerr black hole. This effect is amplified for smaller values of $w$. In the case, where $c<0$ and $\alpha>0$, the retrograde orbit gives rise to a larger deviation than Kerr while in the direct orbit, we can have a larger deviation for small $w$ and smaller deviation for larger $w$.

Finally, we have studied an exact expression for the bending angle in the background of a rotating black hole surrounded by quintessence with EOS parameter $w= -1/3$. We found that the bending angle $\alpha$ of a rotating black hole with quintessence has a lower value compared to the Kerr black hole spacetime in GR when the value of the normalized impact parameter $ b$ lies between $0$ and $0.5$. Also the bending angles of rotating black holes with quintessence coincide with those of the Kerr black holes  for larger values of the normalized impact parameter i.e. $ b > 0.5$. 

\section*{\bigskip Acknowledgments} 
Authors HN and PS thankfully acknowledge the financial support provided by Science and Engineering Research Board (SERB) during the course of this work through grant number EMR/2017/000339. R.G. is supported by Fondecyt project No 1171384. HN is also thankful to IUCAA, Pune (where a part of the work was completed) for support in form of academic visits under its associateship programme. AA acknowledges that this work is based on the research supported in part by the National Research Foundation (NRF) of South Africa (grant numbers 109257 and 112131).AA also acknowledges the hospitality of the High Energy and Astroparticle Physics Group of the Department of Physics of Sultan Qaboos University, where a part of this work was completed.The authors are thankful to Prof. Philippe Jetzer for various useful suggestions during the early stage of this work.

\end{document}